\documentclass[aps,prl,reprint,lengthcheck,twocolumn,floatfix,superscriptaddress,float,nofootinbib,longbibliography,nobalancelastpage]{revtex4-2}

\usepackage{graphicx}
\usepackage{amssymb,amsfonts,amsmath}
\usepackage{braket}
\usepackage{xcolor}
\usepackage{dsfont}
\usepackage[colorlinks=true, linkcolor=black, breaklinks=true, citecolor=blue, urlcolor=red]{hyperref} 

\usepackage{layouts}

\newcommand{\bbZ}{\mathbb{Z}}

\newcommand{\calN}{\mathcal{N}}
\newcommand{\eps}{\varepsilon}
\newcommand{\calR}{\mathcal{R}}
\newcommand{\eff}{\textup{eff}}

\newcommand{\Tr}{\mathrm{Tr\,}}
\newcommand{\1}{\mathds{1}}

\begin{document}
\title{Interacting chiral fermions on the lattice with matrix product operator norms}

\author{Jutho Haegeman}
\affiliation{Department of Physics and Astronomy, Ghent University, Krijgslaan 281, 9000 Gent, Belgium}

\author{Laurens Lootens}
\affiliation{Department of Physics and Astronomy, Ghent University, Krijgslaan 281, 9000 Gent, Belgium}
\affiliation{Department of Applied Mathematics and Theoretical Physics, University of Cambridge,\\ Wilberforce Road, Cambridge, CB3 0WA, United Kingdom}

\author{Quinten Mortier}
\affiliation{Department of Physics and Astronomy, Ghent University, Krijgslaan 281, 9000 Gent, Belgium}

\author{Alexander Stottmeister}
\affiliation{Institut f\"ur Theoretische Physik, Leibniz Universit\"at Hannover, Appelstr. 2, 30167 Hannover, Germany}

\author{Atsushi Ueda}
\affiliation{Department of Physics and Astronomy, Ghent University, Krijgslaan 281, 9000 Gent, Belgium}

\author{Frank Verstraete}
\affiliation{Department of Applied Mathematics and Theoretical Physics, University of Cambridge,\\ Wilberforce Road, Cambridge, CB3 0WA, United Kingdom}
\affiliation{Department of Physics and Astronomy, Ghent University, Krijgslaan 281, 9000 Gent, Belgium}

\begin{abstract}
We develop a Hamiltonian formalism for simulating interacting chiral fermions on the lattice while preserving unitarity and locality and without breaking the chiral symmetry. The fermion doubling problem is circumvented  by constructing a Fock space endowed with a semi-definite norm. When projecting our theory on the the single-particle sector, we recover the framework of Stacey fermions, and we demonstrate that the scaling limit of the free model recovers the chiral fermion field. Technically, we make use of a matrix product operator norm to mimick the boundary of a higher dimensional topological theory.  As a proof of principle, we consider a single Weyl fermion on a periodic ring with Hubbard-type nearest-neighbor interactions and construct a variational generalized DMRG code to demonstrate that the ground state for large system sizes can be determined efficiently. As our tensor network approach does not exhibit any sign problem, we can add a chemical potential and study real-time evolution.
\end{abstract}

\maketitle

\paragraph{Introduction and main results.}
Simulating chiral quantum field theories on the lattice is a central problem in both high energy and condensed matter physics.  The Nielsen-Ninomiya theorem \cite{nielsen1981absence,nielsen1981absence_bis} prohibits the existence of a local symmetry-preserving lattice model with the correct continuum limit due to unphysical low-energy modes (``doublers''), which dynamically couple in the presence of interactions. Any local lattice formulation, e.g., Wilson fermions \cite{wilson1974confinement}, Ginsparg–Wilson fermions \cite{ginsparg1982remnant} or staggered fermions \cite{kogut1975hamiltonian}, either violate an internal or lattice symmetry. The only lattice models exhibiting chiral fermions without breaking any other symmetries are non-local theories \cite{bietenholz1996perfect,drell1976strong,Hernandez:1998et}, which is unsatisfactory. The problem can be mitigated by placing the theory on the edge of a higher-dimensional theory \cite{kaplan1992method,PhysRevLett.132.141603,PhysRevLett.132.141604}, but this leads to much larger simulation costs, and some symmetries are still broken. 

In this Letter, we provide an alternative solution preserving all symmetries without requiring an increase in spacetime dimension by endowing our Hilbert space with a semi-definite metric in the form of a matrix product operator (MPO) \cite{verstraete2004matrix,cirac2021matrix}. Intuitively, this metric mimics the role of the higher-dimensional bulk theory, but its construction is locality preserving. For sake's clarity, we will limit our discussion to the one-dimensional case, but all arguments extend to higher dimensions. 

Our work is based on the ground-breaking paper of Stacey \cite{stacey1982eliminating}, which solved the fermion doubling problem in the single-particle case by making use of the implicit (versus the usual explicit) way of discretizing the differentials in the field equations. We reinterpret this construction from a finite-elements perspective. Following \cite{pacholski2021generalized,beenakker2023tangent}, Stacey's theory amounts to working with non-orthogonal orbitals and leads to a mapping of the stationary solutions of a single-particle Weyl equation 
\begin{equation}
   i \tfrac{\partial\ }{\partial t} \psi(x, t) = -i\tfrac{\partial\ }{\partial x} \psi(x, t)\label{Weyl}
\end{equation}
to a generalized eigenvalue problem of tridiagonal matrices (with periodic boundary conditions on $L$ sites)
\begin{align}
\label{eq:gev}
H|\phi\rangle & =\lambda N|\phi\rangle\;, \\ \nonumber
H_{mn} & \!=\! \tfrac{i}{2}(\delta_{m,n-1}\!-\!\delta_{m,n+1})\,,\!\!\!\! &
N_{mn} & \!=\! \tfrac{1}{4}(\delta_{m,n\mp1}\!+\!2\delta_{m,n})\,,
\end{align}
where $N$ encodes the overlap between the different orbitals. These circulant matrices are are diagonal in the Fourier basis. The generalized eigenvalues are ratios of the eigenvalues of $N$ and $H$ labeled by momenta $k\!=\!\tfrac{2\pi n}{L}$:
\begin{equation}
\label{eq:chiral_disp}
\lambda_k =\tfrac{2\sin(k)}{1+\cos(k)}=2\tan(\tfrac{1}{2}k).
\end{equation}
The essential trait of this solution is that the unwanted zero-modes of $H$ at $ k=\pm \pi$ become poles by virtue of the norm of the associated wavefunction being singular: the metric $N$ has a zero mode breaking the continuity of the dispersion relation (lying at the origin of fermion doubling). 

\begin{figure}
    \centering
    \includegraphics{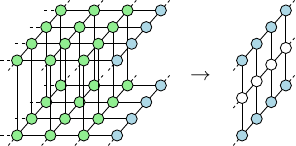}
    \caption{The matrix product operator fixed point of a double layer PEPS (green) induces a matrix product operator norm (white) for matrix product states (blue) living on the virtual degrees of freedom of the PEPS.}
    \label{fig:bulk_boundary}
\end{figure}

In this paper, we extend Stacey's formalism to the many-body case, i.e., we build its second-quantized version. The Fock space is spanned by (unnormalized) vectors
\[ |n_1,n_2,..., n_L\rangle=(a_1^\dag)^{n_1}(a_2^\dag)^{n_2}...(a_L^\dag)^{n_L}|\Omega\rangle\]
with $n_i\in\{0,1\}$, $a_i|\Omega\rangle=0$ and commutation relations
\begin{align}
\label{eq:a_alg}
\{a_i,a_j^\dag\} & = N_{ij},&\{a_i,a_j\} = \{a_i^\dag,a_j^\dag\}&=0
\end{align}
where $N$ is the single-particle norm matrix. This choice reproduces Stacey's results in the single-particle sector. A crucial feature of the usefulness of this Fock space representation is contained in the following fact: the overlaps between different Fock states are given by a matrix product operator $\tilde{N}$ of bond dimension $\chi=4$. This implies that we can easily calculate norms (and expectation values) of matrix product state wavefunctions of the form
\begin{align}
\label{eq:mps_wave}
|\psi\rangle=\sum_{n_1n_2\cdots}\underbrace{\Tr\left(A_1^{n_1}A_2^{n_2}... A_L^{n_L}\right)}_{|\phi\rangle}|n_1,n_2,...,n_L\rangle
\end{align}
by contracting the tensor network $\langle\phi|\tilde{N}|\phi\rangle$. The physical meaning of the semidefinite norm becomes transparent by imagining that the chiral theory lives on the edge of a (symmetry protected) topological model in 2+1 dimensions, and more specifically of a gapped bulk theory, represented by a projected entangled-pair state (PEPS) \cite{verstraete2004renormalization,cirac2021matrix}. The edge modes can be represented by the dangling indices of the PEPS on the border \cite{yang2014edge}, and the norm of such states is determined by the entanglement spectrum of the bulk theory (see Fig.~\ref{fig:bulk_boundary}). In \cite{wahl2013projected,wahl2014symmetries}, it was demonstrated that entanglement Hamiltonians of chiral PEPS exhibit zero-modes similar to the ones represented here. Nevertheless, our simulations do not require such a higher-dimensional theory, which differentiates our approach from the one in \cite{PhysRevLett.132.141603,PhysRevLett.132.141604}.

\begin{figure}
    \centering
    \includegraphics{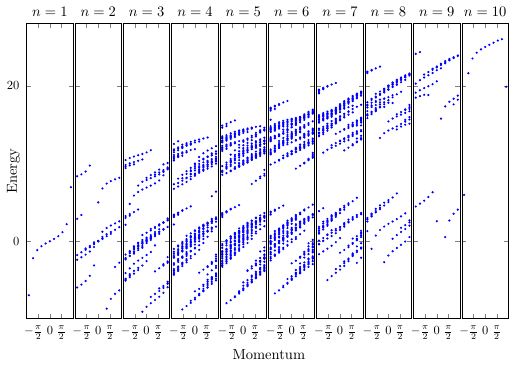}
    \caption{The full generalized spectrum of a chiral fermion in the presence of a Hubbard-type interaction \eqref{H} on 11 sites with $n$ particles with $J=U=1$. In the single-particle sector, we recover the correct single-particle spectrum.}
    \label{fig:L_11_U_1}
\end{figure}

The Hamiltonian, just as any other observable, is expressed in terms of conjugate creation/annihilation operators $b_i$. We will simulate the following Hamiltonian:
\begin{equation}
\hat{H}\!=\!J\sum_{n} \left(i\hspace{1mm} b_n^\dagger b_{n+1} + h.c.\right) \!+\! U\sum_n b_n^\dagger b_{n+1}^\dagger b_{n+1}b_n,\label{H}
\end{equation}
The $b_i$ also annihilate the vacuum $|\Omega\rangle$ and satisfy the canonical (anti)commutation relations
\begin{align}
\label{eq:ba_alg}
\{b_i,a_j^\dag\} & =\delta_{ij}.
\end{align}
The knowledge of this (anti)commutation rule is sufficient to calculate expectation values of local observables efficiently using tensor networks. Indeed, the Hamiltonian $\hat{H}$ can be retracted onto $\ket{\phi}$, which we denote as $\tilde{H}$. This retraction preserves hermiticity such that $\tilde{N}$ and $\tilde{H}$ give rise to a Hermitian generalized eigenvalue problem with a real spectrum. In the single-particle sector, it reduces to Stacey's formulation \eqref{eq:gev}. But our constructions allows to target the interacting many-body problem, as illustrated in Fig.~\ref{fig:L_11_U_1}, showing the full generalized spectrum of the above Hamiltonian on a lattice with $L=11$ sites, together with the associated momenta. In the case of non-interacting fermions ($U=0$), the ground state is simply obtained by filling the Fermi sea yielding a ground state energy equal to the sum of negative single-particle generalized eigenvalues.
The dynamics of the system is governed by a generalized Schrödinger equation
\begin{equation}
    i\tilde{N}\tfrac{\partial\ }{\partial t}|\phi(t)\rangle=\tilde{H}|\phi(t)\rangle
\end{equation}
and both $\langle\phi(t)|\tilde{N}|\phi(t)\rangle$ and $\langle\phi(t)|\tilde{H}|\phi(t)\rangle$ are  constants of motion under these dynamics.
Since $\tilde H$ and $\tilde N$ are $U(1)$ invariant, we can use the gauging map for quantum states \cite{haegeman2015gauging,lootens2023dualities,lootens2024dualities} to gauge the $U(1)$ symmetry of this model. As we are dealing with a single right-moving Weyl fermion this $U(1)$ symmetry is well known to be anomalous in the continuum quantum field theory prohibiting its promotion to a gauge symmetry. We expect this anomaly to manifest itself in our lattice model, for instance through the Lieb-Schultz-Mattis theorem \cite{lieb1961two} as a mixed anomaly between the $U(1)$ and translation symmetry \cite{PhysRevB.96.195105}.

\paragraph{Action principle and quantization.}
Recently, Stacey's formalism was used for discretizing both space and time in a path integral formalism, resulting in a local Lagrangian for a helical Luttinger liquid with Hubbard interaction, amenable to Monte Carlo sampling at half-filling \cite{zakharov2024helical}. Here, instead, we keep time continuous in order to motivate the second quantization and Fock space construction, as well as the resulting generalized Schrödinger equation. Furthermore, we reinterpret the spatial lattice as resulting from applying a finite elements procedure using a set of non-orthogonal basis functions to approximate the continuum problem. We start from action for which the Euler-Lagrange equation gives rise to the Weyl equation (\ref{Weyl})
\begin{align*}
    S[\psi^\ast,\psi] & \!=\!\int_{t_0}^{t_1} \!\!\mathrm{d}t\int\!\!\mathrm{d} x\,\Big( \tfrac{i}{2}\psi(x,t)^\ast\tfrac{\partial}{\partial t}\psi(x,t)\\
    &\hspace{-1cm} -\!\tfrac{i}{2}\!\left(\tfrac{\partial}{\partial t}\psi^\ast(x,t)\right)\psi(x,t)\!-\!\psi(x,t)^\ast\!\left(\!-i \tfrac{\partial }{\partial x}\right) \psi(x,t)\Big)\;.
\end{align*}
Here, the time derivative has been written in a manifestly Hermitian form, whereas the spatial derivative is Hermitian by virtue of the domain and spatial boundary conditions of $\psi$ at any instance of time. We express $\psi(x,t)$ with respect to a set of functions $\{\varphi_n(x), n\in \bbZ\}$ as
\begin{equation}
\label{eq:non_expansion}
\psi(x,t) = \sum_{n} b_n(t) \varphi_n(x)
\end{equation}
where the functions are not orthonormal but satisfy
\begin{align}
\label{eq:unit_scale}
\langle\varphi_{m}|\varphi_{n}\rangle & \!=\!N_{mn}\;, & 
\langle\varphi_{m}|(\!-\!i\tfrac{\partial}{\partial x})|\varphi_{n}\rangle & \!=\!H_{mn}\;. 
\end{align}
An admissible set of functions that satisfy these identities in a distributional sense is given by $\varphi_{n}(x)\!=\!\tfrac{1}{\sqrt{2}}\chi_{(-1,+1)}(2x\!-\!n)$, where $\chi_{(-1,+1)}(x)$ is the indicator function of the open interval $(-1,+1)$.
Inserting \eqref{eq:non_expansion} into the action gives rise to $S[b^\ast,b]\!=\!\int_{t_0}^{t_1}\mathrm{d}t\,L(b,\dot{b},b^\ast,\dot{b}^\ast)$ with
\begin{align*}
L & \!=\!\sum_{m,n}\!\left\{\!\tfrac{i}{2}\!\left[ b_m^\ast N_{mn} \dot{b}_n\!-\!\dot{b}_m^\ast N_{mn} b_n\right]\!-\!b_m^\ast H_{mn} b_n \!\right\}.
\end{align*}
The canonical conjugate variables to $\{b_{n}\}$ are given by
\begin{align}
\label{eq:conj_mom}
p_{n} = \frac{\partial L}{\partial \dot{b}_{n}} & = -ib_{m}^{*}N_{mn}
\end{align}
We quantize the theory by imposing canonical anti-commutation relations $\{p_{m} , b_{n}\} = -i \delta_{m,n}$
thereby obtaining \eqref{eq:a_alg} and \eqref{eq:ba_alg} by defining $a_n\!=\!b_m N_{mn}$ and, thus, $a_n^\dagger\!=\!b_m^\dagger N_{mn}$\footnote{Strictly speaking, all steps only hold true in this form if $N$ is non-degenerate, which can be achieved by choosing appropriate boundary conditions on finite lattices. This does not pose any problem for the continuum and thermodynamic limit as long as $0$ belongs to the continuous spectrum of $N$. Otherwise, Dirac's algorithm for constrained systems needs to be applied.}.
The second-quantized Hamiltonian is given by
\begin{align}
\label{eq:latt_ham}
\hat{H} & = \sum_{m,n} b_m^\dagger H_{mn} b_n = \sum_{m,n} a_m^\dagger (N^{-1} H N^{-1} )_{mn} a_n
\end{align}
to which we can add interaction terms, as in \eqref{H}. The free Heisenberg equations of motion are
\begin{align*}
\dot{a}_k^\dagger\!=\!i[\hat{H}, a_k^\dagger]\!=\!i\sum_{m} b_m^\dagger H_{mk}\!=\!i\sum_{m} a_m^\dagger (N^{-1} H)_{mk},
\end{align*}
which entail local commutators with the Hamiltonian for observables built from $b^{\dag}_{m}N_{mn}$ and $b_{m}N_{mn}$, e.g.,
\begin{align*}
[\hat{H}, b^{\dag}_{m}N_{mn}b_{k}N_{kl}] & = b^{\dag}_{m}H_{mn}b_{k}N_{kl}-b^{\dag}_{m}N_{mn}b_{k}H_{kl}.
\end{align*}
We expect this to be relevant to prove Lieb-Robinson-type bounds in our model.
We can now build a Fock space using the creation operators $a_m^\dagger$ acting on a vacuum $\ket{\Omega}$ with $a_n\ket{\Omega}=0 = b_k \ket{\Omega}$. We expand a many-body state as
\begin{equation}
\ket{\Psi} = \sum_{\{n_i\}} \phi_{n_{1} n_{2}...}\,(a_{1}^\dag)^{n_1}(a_{2}^\dag)^{n_2}...\ket{\Omega}\label{eq:psi}
\end{equation}
Denoting $\phi_{\vec{n}}\!=\!\phi_{n_{1} n_{2}...}$ and $\ket{\vec{n}}\!=\!(a_{1}^\dag)^{n_1}(a_{2}^\dag)^{n_2}... \ket{\Omega}$, the time-evolution action principle $S[\phi^\ast\!, \phi]\!=\!\int_{t_0}^{t_1}\!\mathrm{d} t L(t)$ with 
\begin{align*}
L(t) & =\sum_{\vec{n},\vec{n}'}  \left[ \tfrac{i}{2} \phi_{\vec{n}}^\ast \tilde{N}_{\vec{n},\vec{n}\prime} \dot{\phi}_{\vec{n}'}+ h.c. \right] - \phi_{\vec{n}}^\ast \tilde{H}_{\vec{n},\vec{n}'} \phi_{\vec{n}'}\nonumber
\end{align*}
gives rise to the equation of motion
\begin{align*}
i \sum_{\vec{n}'}\tilde{N}_{\vec{n},\vec{n}'} \dot{\phi}_{\vec{n}'} & = \sum_{\vec{n}'}\tilde{H}_{\vec{n},\vec{n}'} \phi_{\vec{n}'}
\end{align*}
with $\tilde{N}_{\vec{n},\vec{n}'} = \braket{\vec{n}|\vec{n}'}$ and $\tilde{H}_{\vec{n},\vec{n}'} = \braket{\vec{n}|\hat{H}|\vec{n}'}$ the retraction of the identity and Hamiltonian as introduced above. If the set of coefficients $\phi$, previously referred to as $\ket{\phi}$, is also constrained to a particular form such as MPS, we can still use the same action by inserting such parameterization. This gives rise to the Dirac-Frenkel time-dependent variational principle (TDVP), where the retraction of the norm matrix onto the tangent space of the variational manifold will appear. Importantly, we will not have to invert the norm matrix $\tilde{N}$ in the full Hilbert space, but only its restriction onto the tangent space.

\paragraph{Continuum limit.}
\label{sec:cont_lim}
The continuum formulation of a (free) chiral fermion field describing right-moving particles can be recovered by a scaling limit procedure \cite{stottmeister2020oar, osborne2023conformal}: We start with the construction in finite volume resolved at various length scales $\Lambda_{\eps}\!=\!\eps\{1,...,L\}\!\subset\!\eps\bbZ$ ($\eps L\!=\!\textup{const.}$) as the thermodynamic limit can be achieved consecutively \cite{morinelli2021scaling_limits_wavelets, Osborne2023IsingFP}.
In \eqref{eq:unit_scale}, the model is defined at unit scale ($\eps=1$). It can be transferred to $\Lambda_{\eps}$ by rescaling: $\varphi^{(\eps)}_{n}(x)\!=\!\tfrac{1}{\sqrt{\eps}}\varphi_{n}(\eps^{-1}x)$.
To extract the low-energy physics of the lattice model, we focus on the many-body ground-state sector associated with \eqref{eq:latt_ham} at a given scale $\eps$, i.e., by filling up all the negative-energy modes according to the dispersion relation in \eqref{eq:chiral_disp}: $\ket{\textup{vac}_{\eps}}\!\propto\!\prod_{k<0}a^{\dag}_{k}\ket{\Omega_{\eps}}$. The scaling limit is constructed by defining an ascending superoperator $\calR$ identifying normal-ordered Wick monomials in $a, a^{\dag}$ at different scales (following Wilson's triangle \cite{wilson1975kondo,stottmeister2020oar}). This allows us to obtain the convergence correlation functions $\bra{\textup{vac}_{\eps}}a^{\dag}...a...\ket{\textup{vac}_{\eps}}$ at finite scales to the vacuum correlation functions $\bra{\textup{vac}_{0}}c^{\dag}...c...\ket{\textup{vac}_{0}}$ of chiral fermion field $\psi$ in the continuum (with annihilation/creation operators $c,c^{\dag}$).
The convergence relies on the observation that the metric at scale $\eps$ converges to the identity in the scaling limit, i.e., $N_{\eps}\stackrel{\eps\to 0}{\rightarrow}\1$. This entails that (\ref{eq:a_alg}) reproduces the canonical anticommutation relations in the scaling limit. At the same time, the scale-dependent dispersion relation $\lambda^{(\eps)}_{k}\!=\!2\eps^{-1}\tan(\tfrac{1}{2}\eps k)$ recovers the massless, relativistic dispersion relation $\lambda^{(0)}_{k}\!=\!k$ for $\eps\to 0$. We will provide further details on the scaling limit construction elsewhere.

\paragraph{MPO representations.}
The (semi)definite norm matrix
\begin{align*}
\tilde{N}_{n_1...n_L; n_1'...n_L'}\!=\!\braket{\Omega| (a_L)^{n_L}...(a_1)^{n_1}(a_1^\dagger)^{n_1'}...(a_L^\dagger)^{n_L'}|\Omega}
\end{align*}
has the form of a uniform matrix product operator (MPO) with bond dimension $4$ and local tensor $\calN$. To see this, note that a particle created at the site $i$ can be annihilated at the same and adjacent sites.
The MPO tensor $\calN$ has the following non-zero entries (where we use the convention that the indices come in the order left, up, right, down; and that value 2 for the up and down --i.e., physical-- indices encodes a fermion present, while 1 indicates no fermion):
\begin{align*}
    \calN_{1111} &= 1, \quad \calN_{1212} = N_{i,i}\\
    \calN_{1221} &= \calN_{2112} = \calN_{1132} = \calN_{3211} =\sqrt{N_{i,i+1}}\\
    \calN_{1242} &= -\calN_{4212} =\calN_{2222} = \calN_{3232} = N_{i,i+1}
\end{align*}
The MPO has different boundary conditions for the even and odd particle sectors. While no boundary term is needed for the odd particle sector, the even particle sector requires the introduction of a twist with a diagonal matrix with elements $B_{ii}=1$ for $i=1,4$ and $B_{ii}=-1$ for $i=2,3$. This can be checked by mapping $a_i$ to a spin operator. These boundary conditions originate from the Jordan-Wigner transformation of fermions, leading to the extra negative sign in the even particle number sector~\cite{lieb1961two,o2024local}. Alternatively, interpreting $\mathcal{N}$ as a fermionic tensor (using the formalism of \cite{mortier2024fermionic}), with odd fermion parity for value 2 of the physical (up and down) indices and for values $2$ and $3$ of the virtual (left and right) indices, no additional boundary matrix is required.

The Hamiltonian can be constructed from this MPO by acting on it with spin creation and/or annihilation operators $\sigma^+=|1\rangle\langle 0|$, $\sigma^-=|0\rangle\langle 1|$. For example, the kinetic term which acts between sites $k$ and $k+1$ would be represented as $i\sigma^+_{\tilde{n}_k,n_k}\tilde N^{n_1\cdots n_L}_{n'_1\cdots n'_L}\sigma^-_{\tilde{n}'_{k+1}n'_{k+1}} +$ hermitian conjugate.

\paragraph{Interacting ground state optimization.}
Given the fact that both $\tilde{H}$ and $\tilde{N}$ have the form of an MPO and that the variational principle still works for generalized eigenvalue problems whenever $\tilde{N}$ is positive semidefinite (as long as the zero modes for $\tilde{N}$ are also zero-modes for $\tilde{H}$)
\[\lambda_{\min}=\min_{\phi}\frac{\langle\phi|\tilde{H}|\phi\rangle}{\langle\phi|\tilde{N}|\phi\rangle}\]
we can, in principle, use standard tensor network techniques to optimize the MPS \cite{white1992density,verstraete2008matrix}, which allows to simulate interacting spin chains at a cost that only scales linearly in the system size as long as the ground state satisfies an area law for the entanglement entropy  \cite{verstraete2006matrix,hastings2007area}. Crucially, DMRG does not exhibit any sign problem, even away from halve-filling. For the non-interacting model, the filled Fermi sea results in a diverging energy density $E/L$ in the thermodynamic limit, which we expect to persist in the interacting Hamiltonian \eqref{eq:latt_ham}. This prevents a naive implementation of the infinite lattice, but the ground state and its energy on finite periodic rings with an odd number of sites is well defined.
The filled Fermi sea of the non-interacting Hamiltonian gives rise to a ground state momentum that cannot equal zero or $\pi$. Therefore, we have to use a general non-translational MPS ansatz on a ring. Similar to the density matrix renormalization group (DMRG) algorithm for systems with periodic boundary conditions \cite{verstraete2004dmrg}, the MPS can be optimized locally by finding the optimal tensor. More precisely, we compute an effective Hamiltonian $H_{\eff}(i)$ for site-$i$ so that $\langle\psi(A)|\tilde{H}|\psi(A)\rangle = A_i^\dagger \tilde{H}_{\eff}(i) A_i$. Similarly, we obtain $\tilde N_{\eff}(i)$ as $\langle\psi(A)|\tilde{N}|\psi(A)\rangle = A_i^\dagger \tilde N_{\eff}(i) A_i$. Then, the optimal tensor $A_i$ can be found by solving
\begin{equation*}
\tilde H_{\eff}(i) A_i = \lambda_{\min} \tilde N_{\eff}(i)A_i,
\end{equation*}
where $\lambda_{\min}$ is the smallest (generalised) eigenvalue corresponding to the ground state. It is important to project out the null space before inverting $\tilde{N}_{\text{eff}}$. After optimizing every site in an iterative manner, we observe that the energy converges quickly, as in conventional DRMG simulations.
\begin{figure}[tb]
    \centering
    \includegraphics[width=86mm]{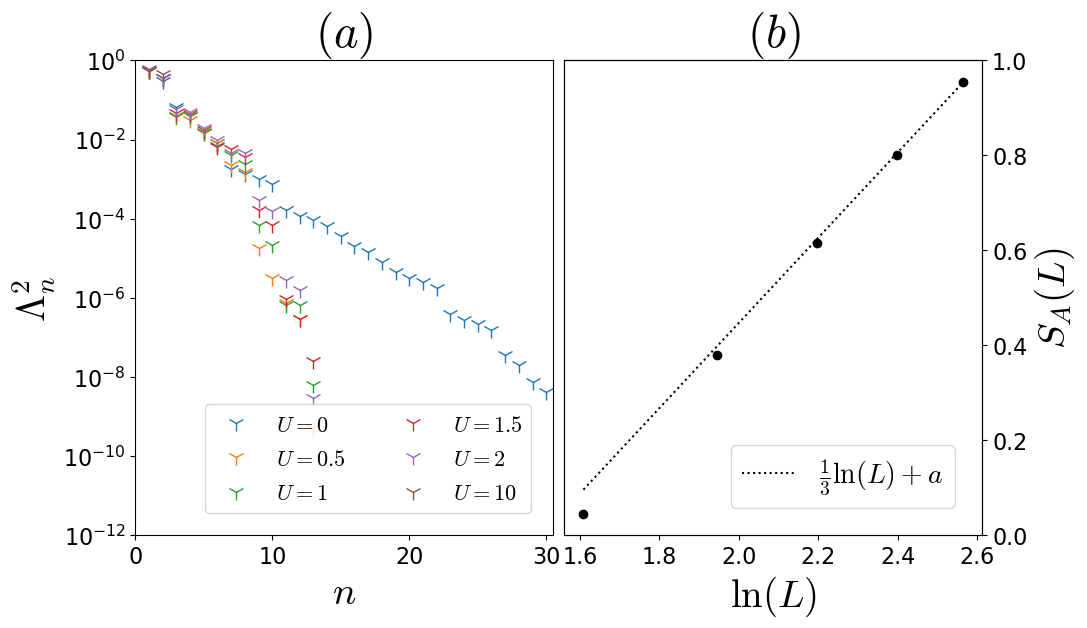}
    \caption{$(a)$ The spectrum of the reduced density matrix obtained from exact diagonalization of the Hamiltonian \eqref{H} on 11 sites. The reduced density matrix is obtained by tracing out the degrees of freedom on the first five sites. $(b)$ The entanglement entropy of the free Hamiltonian ($U=0$), obtained by dividing the lattice by half on 5, 7, 9, 11, and 13 sites. The scaling is consistent with the chiral central charge $c=1$ of complex fermionic modes.}
    \label{fig:schmidt}
\end{figure}
We simulate the interacting Hamiltonian (\ref{H}) with $U/J=1$ and observe that the Schmidt spectrum of the half-chain decays exponentially. This clearly suggests that DMRG will work equally well for the generalized eigenvalue case as for the normal case. For the case of an odd number of particles and length $L=11$; the ground state energies obtained for bond dimension $\chi=(4,6,8,10,12)$ are respectively equal to (-9.084089, -9.086027, -9.086092, -9.086096,-9.086097), which shows the fast convergence to $-9.086098$: the value obtained from exact diagonalization\footnote{We use $N_{i,i}=2, N_{i,i+1}=1$ in the following calculations. }. The success of our DMRG algorithm is corroborated by the hierarchical entanglement structure of many-body ground states in one dimension. In Fig.~\ref{fig:schmidt}, we demonstrate the spectrum of the reduced density matrix computed from exactly diagonalizing the Hamiltonian~\eqref{H} on 11 sites. As is observed in conventional MPS simulations, the Schmidt values decay exponentially and thereby allow for an efficient simulation using MPS with finite bond dimensions. $U=0$ is the gapless regime where we observe a slower decay rate. Still, the spectrum of $U>0$ can be well approximated with a few leading values.  For $L=51$ and $U/J=1$, it is, of course, completely impossible to diagonalize the Hamiltonian with exact diagonalization. We obtain an energy of $-32.889236$ with a DMRG simulation with $\chi=4$.
In analogy to standard MPS techniques, developing finite-size and/or entanglement scaling methods~\cite{PhysRevB.78.024410,PhysRevLett.102.255701,PhysRevB.86.075117,PhysRevB.91.035120,sherman2023universality,PhysRevB.108.024413,PhysRevLett.132.086503} to extrapolate those results is significant to further improve accuracy in the future.

\paragraph{Discussion and outlook.}
Our results indicate that we can build lattice models of interacting chiral fermions, and efficiently target their ground state using ED \cite{doi:10.1137/S1064827500382579} or MPS techniques. Developing finite-size and/or entanglement scaling methods~\cite{PhysRevB.78.024410,PhysRevLett.102.255701,PhysRevB.86.075117,PhysRevB.91.035120,sherman2023universality,PhysRevB.108.024413,PhysRevLett.132.086503} to extrapolate those results is significant to further improve accuracy in the future. It is also important that our method can be extended to the higher-dimensional case by using Projected Entangled Pair States (PEPS) \cite{verstraete2004renormalization,cirac2021matrix}; indeed, it can readily be seen that the norm matrix in this case can efficiently be represented as a PEPO\cite{InPreparation}. Another question is to construct PEPS wavefunctions with the MPO $\tilde{N}$ as the leading eigenvector of the transfer matrix.

We can also generalize  MPS techniques to extract the interacting excitation spectrum \cite{pirvu2012matrix,vanderstraeten2019tangent} or to study dynamics. For the latter, the TDVP equation \cite{haegeman2011time} seems to provide the best starting point, as the Hamiltonian $\tilde{H}$ acting on the MPS includes the norm matrix and does not admit a simple Trotter decomposition. While the effective norm matrix in the MPS tangent space (a.k.a.~the metric) can no longer be made equal to the identity matrix using MPS gauge manipulations, the resulting equations can still be integrated efficiently by inverting the Gram matrix using a conjugate gradient scheme\cite{InPreparation}.  

These algorithmic considerations are closely related to the conceptual questions regarding the locality of the generalized Schrödinger equation, which we will address in a forthcoming paper. What is the correct generalization of the Lieb-Robinson bound in this nonorthogonal quantum world, and is satisfied by a Hamiltonian that is local in $b$ and $b^\dagger$? Can an area law for the (Renyi) entanglement entropy in the generalized eigenvalue setting \cite{verstraete2006matrix,hastings2007area} be proven?  It is also an interesting question to compare our Hamiltonian approach to the path integral approach of \cite{zakharov2024helical}, where both time and space were discretized with Stacey's method. Another outstanding question is how to simulate our generalized Hamiltonian system on a quantum simulator or quantum computer. It is not evident to do this without coupling the system of interest to a bulk theory or by introducing long-range interactions in the form of $\tilde{N}^{-1/2}\tilde{H}\tilde{N}^{-1/2}$. 

\medskip\noindent
\textbf{Acknowledgments:} \emph{We thank Carlo Beenakker for introducing us to the concept of tangent fermions and Nick Bultinck, Karel Van Acoleyen and David Tong for helpful discussions.  This work has received funding from EOS (grant No. 40007526), IBOF (grant No. IBOF23/064), BOF-GOA (grant No. BOF23/GOA/021), FWO (grant No. 12AUN24N and GOE1520N) and the MWK Lower Saxony via the Stay Inspired Program (Grant ID: 15-76251-2-Stay-9/22-16583/2022). }

\bibliography{references.bib}

\end{document}